\documentclass[12pt,preprint]{aastex}
\usepackage{amsmath,amssymb,amsfonts,longtable,rotating}
\usepackage{epsfig}                   
\usepackage{epstopdf}                   
\usepackage{multirow}
\usepackage{natbib}
\usepackage{txfonts}                  
\usepackage{graphicx}
\usepackage{lscape}
\usepackage{rotating}

\shorttitle{}
\shortauthors{}
\begin{document}

\newcommand{\ctan}{$^{13}$C($\alpha$,n)$^{16}$O }
\newcommand{\nean}{$^{22}$Ne($\alpha$,n)$^{25}$Mg }
\newcommand{\neag}{$^{22}$Ne($\alpha$,$\gamma$)$^{25}$Mg }
\newcommand{\cdpg}{$^{12}$C(p,$\gamma$)$^{13}$N }
\newcommand{\nqnp}{$^{14}$N(n,p)$^{14}$C }
\newcommand{\cd}{$^{12}$C}
\newcommand{\ct}{$^{13}$C }
\newcommand{\nq}{$^{14}$N }
\newcommand{\odo}{$_\odot$~}
\newcommand{\s}{{\it s}}

\title{Constraints of the physics of low-mass AGB stars from CH and CEMP stars}

\author{S. Cristallo\altaffilmark{1,2}
\affil{1 - INAF - Osservatorio Astronomico di Teramo, 64100 Italy}
\affil{2 - INFN- Sezione di Perugia, Italy}}

\and

\author{D. Karinkuzhi\altaffilmark{3}
\affil{3 - Indian Institute of Astrophysics, Koramangala,
Bangalore 560034}}

\and

\author{A. Goswami\altaffilmark{3},
\affil{3 - Indian Institute of Astrophysics, Koramangala,
Bangalore 560034}}

\and

\author{L. Piersanti\altaffilmark{1,2},
\affil{1 - INAF - Osservatorio Astronomico di Teramo, 64100 Italy}
\affil{2 - INFN- Sezione di Perugia, Italy}}

\and

\author{D. Gobrecht\altaffilmark{1,2}
\affil{1 - INAF - Osservatorio Astronomico di Teramo, 64100 Italy}
\affil{2 - INFN- Sezione di Perugia, Italy }}



\begin{abstract}
We analyze a set of  published elemental abundances from a sample
of CH stars which are based on high resolution spectral analysis
of ELODIE and  SUBARU/HDS spectra. All the elemental abundances
were derived from local thermodynamic equilibrium analysis using
model atmospheres, and thus, they represent the largest
homogeneous abundance data available for CH stars up to date. For
this reason, we can use the set to constrain the physics and the
nucleosynthesis occurring in low mass AGB stars. CH stars have
been polluted in the past from an already extinct AGB companion
and thus show s-process enriched surfaces. We discuss the effects
induced on the surface AGB s-process distributions by different
prescriptions for convection and rotation. Our reference
theoretical FRUITY set fits only part of the observations.
Moreover, the s-process observational spread for a fixed
metallicity cannot be reproduced. At [Fe/H]$>$-1, a good fit is
found when rotation and a different treatment of the inner border
of the convective envelope are simultaneously taken into account.
In order to increase the statistics at low metallicities, we
include in our analysis a selected number of CEMP stars and,
therefore, we compute additional AGB models down to [Fe/H]=-2.85.
Our theoretical models are unable to attain the large [hs/ls]
ratios characterizing the surfaces of those objects. We speculate
on the reasons for such a discrepancy, discussing the possibility
that the observed distribution is a result of a proton mixing
episode leading to a very high neutron density (the so-called {\it
i}-process).

\end{abstract}


\section{Introduction}
CH stars are Main Sequence or Giant Stars showing a large variety
of nuclear species. On their surfaces, in fact, the abundances of
numerous heavy elements, normally attributed to the slow neutron
capture process (the s-process), have been detected. Those
elements cannot be produced {\it in situ} and, therefore, must
have been accreted by already evolved companions. Best candidates
are Asymptotic Giant Branch (AGB) stars, which are responsible for
the synthesis of the Main component of the s-process
\citep{ga98,busso99}. Therefore, an observational study of CH
stars could contribute to improve our knowledge of the physical
processes at work in the internal layers of AGB stars. The
structure of an AGB consists of an inert partially degenerate CO
core, surrounded by a He-shell, separated from a H-shell by a thin
stellar layer (named He-intershell) and by a cool and extremely
expanded convective envelope. During the AGB phase, the surface
luminosity is mainly sustained by the H-burning shell. However,
when its ashes reach high enough temperature and density,
3$\alpha$ reactions trigger a violent thermal runaway (Thermal
Pulse, TP). The sudden energy release cannot be transported away
radiatively and, therefore, convection mixes the whole
He-intershell. As a consequence, the above layers (including the
H-shell) expand and cool down. When convection ceases, the entropy
barrier normally provided by the H-shell cannot prevent the
penetration of the convective envelope in the underlying stellar
region and, therefore, freshly synthesized nuclei are carried to
the surface (Third Dredge Up, TDU). Later, the H-shell switches on
again and this cycle repeats until the envelope is almost entirely
lost via strong stellar winds. For a detailed description of AGB
evolution
see \cite{he05}, \cite{stra06} and \cite{kakka14}.\\
The main neutron source in AGB stars is the \ctan reaction, which
burns radiatively between two TPs \citep{stra95}. The \ct
reservoir needed to reproduce observed distributions is stored in
a tiny region below the formal border of the convective envelope
during TDU episodes. In that layer, a mechanism able to shape or
stretch the sharp proton profile left by convection is required.
Later, when the temperature rises again, those protons are
captured by the abundant $^{12}$C, leading to the formation of the
so-called $^{13}$C-pocket. The physics of such a mechanism is a
highly debated subject and many hypotheses have been postulated:
gravity waves plus Kelvin-Helmholtz instabilities \citep{battino},
magnetic fields \citep{trippa2}, stochastic mixing events
\citep{bisterzo2015} and ballistic plumes penetration
\citep{cri09}. Moreover, the possibility that other mechanisms,
such as rotation, stretch the recently formed \ct profile
at a later time cannot be excluded \citep{pi13}.\\
In this paper we compare a sample of CH stars, spanning over a
large range of metallicities, with our AGB theoretical models.
Rather than focusing on the comparison between each star with the
corresponding model, we aim to verify if the treatment of the
various physical processes in our models is accurate enough to
allow the reproduction of the whole CH sample. In particular, we
want to test if our theoretical recipe to model the formation and
evolution of the $^{13}$C-pocket in AGB stars is reliable or not.
In Section \ref{obser} we present the observations and we discuss
the errors affecting observational data. In Section \ref{modtheo}
we describe our reference theoretical scenario used to interpret
observations and in Section \ref{refere} we compare it to data. In
Sections \ref{rotate} and \ref{convetta} we explore the
consequences deriving from different prescriptions for convection
and rotation on the theoretical interpretation of data. In Section
\ref{lowz} we extend our analysis to Carbon Enhanced Metal Poor
(CEMP) stars. Finally, we deduce our conclusions in Section
\ref{conclu}.

\section{Observations} \label{obser}
Program stars are selected from the CH star catalogue of
\cite{bartkevicius1996} and HES survey of \cite{christlieb2001}.
The basic data of these objects along with the temperatures and
radial velocities are listed in \cite
{goswami2006,goswami2010,goswami2016,kari2014,kari2015}. The
spectra for most of these objects are taken from the  ELODIE
archive \citep{moultaka2004}.  We have considered only those CH
stars for which high resolution spectra are available in the
archive with S/N ratio $>$ 20. ELODIE is a cross dispersed echelle
spectrograph used at the 1.93 m telescope of Observatoire de Haute
Provence (OHP). Details about the spectrograph and reduction
procedures are given in \cite{baranne1996}. The spectra recorded
in a single exposure as 67 orders on a 1K CCD have a resolution of
R = 42000. The wavelength range spans from  3900 \AA\, to 6800
\AA\,. High-resolution spectra (R$\sim$50000) for a few objects
HD~26, HD~5223, HD~198269, HD~224959, HD~209621, HE~1305+0007 and
HE~1152-0355 were obtained using the High Dispersion Spectrograph
(HDS) attached to the 8.2m Subaru Telescope \citep{noguchi2002}.
The observed bandpass ran from about $4020$\,{\AA} to
$6775$\,{\AA}, with a gap of about $75$\,{\AA}, from $5335$\,{\AA}
to $5410$\,{\AA}, due to the physical spacing of the CCD
detectors.

The program stars are divided into three groups  i.e., Group I
(objects that are  known binaries), Group II (objects that are
known to be radial velocity variables), and Group III (objects
with no information on radial velocity variations). Such a
classification helps in comparing the abundance patterns of the
Group III objects with their counterparts observed in Group I and
Group II objects. In particular, it would be interesting to check
if any signature derived from the surface chemical composition of
Group III objects can be related to binarity. The latter is an
essential requirement to explain the surface chemical composition
of CH stars characterized by enhancements of carbon and
neutron-capture elements.

\subsection{Stellar atmospheric parameters and elemental abundances}
Detailed description of the procedures  used for the calculation
of the basic atmospheric parameters and elemental abundances along
with the results are given in
\cite{goswami2006,goswami2010,goswami2016,kari2014,kari2015}. It
is important to notice that, up to date, our sample is the largest
one determined with the same analysis technique and atmosphere
models. Here we briefly discuss the procedures that are relevant
for the present work.
 The atmospheric parameters for deriving the chemical abundances
are determined using a set of unblended Fe I and Fe II lines.
Elemental abundances are derived from measuring the equivalent
widths for a few elements whenever we could measure more than one
line for that element. We have done  spectrum synthesis
calculation for the elements that are known to be affected by
hyperfine splitting. These calculations are performed by using a
recent version of MOOG \citep{sneden1973} assuming Local
Thermodynamic Equilibrium (LTE). A detailed error calculation for
the elemental abundances presented in tables 1 - 3 is discussed in
the next Section.

\subsection{Error Analysis}

We have calculated the errors in our abundance results as
described in \cite{ryan1996}. The errors in the elemental
abundances have contributions from the uncertainty in measurement
of the atmospheric parameters and equivalent widths. The errors in
the equivalent width also have an effect on the measurement of
surface gravity and micro-turbulent velocities. Inaccurate
gf-values also contribute to the errors in the elemental
abundances. For the elements whose abundances have been determined
using spectrum synthesis method, the error is uniformly taken as
0.2 dex. This estimate takes into account the fitting error (0.1
dex) and the error in gf-values (0.1 dex). This error evaluation
holds for Sc, V, Mn, Ba, La and Eu. Moreover, we could see a clear
variation in the abundance profiles when the synthetic spectra are
plotted with $\pm$ 0.2 dex.   For a few objects we have calculated
the elemental abundances using a single line. In that case the
error is expected to be determined from the measurement of
atmospheric parameters only. To find the minimum error for each
star, we have considered the respective standard deviation of the
iron abundances for each object (approximately 0.1 dex), along
with the uncertainties in temperature, micro-turbulent velocities
and surface gravities. We have assumed an error of 100 K  in the
effective temperature, which leads to an error of 0.1 dex in
abundances. Similarly, we estimate 0.03 dex from gravity
variations and 0.06 dex from micro-turbulence changes
(corresponding to a change of 0.5 km/s). These values are
typically accepted as the minimum error in giants and sub-giants
\citep{ryan1996,aoki2007}. The total minimum error is calculated
using Eq. \ref{equa} and the value is found to be 0.12. This value
is assumed as the error for elements for which a single line is
used for the abundance calculation.
\begin{equation} \label{equa}
E_r = \sqrt{{E_{r1}}^2 + {E_{r2}}^2 + {E_{r3}}^2 + {E_{r4}}^2+
{E_{r5}}^2......+ {E_{rn}}^2}
\end{equation}

When more than one line is used  for the abundance calculation,
the standard deviation of the abundances derived using individual
lines is taken as the error. In Tables 1, 2 and 3, we present the
errors for the elements for which abundances are determined by
equivalent width measurements of two or more lines. In Table 4, we
have presented the
[ls/Fe]\footnote{[ls/Fe]=([Sr/Fe]+[Y/Fe]+[Zr/Fe])/3},
[hs/Fe]\footnote{[hs/Fe]=([Ba/Fe]+[La/Fe]+[Nd/Fe]+[Ce/Fe]+[Sm/Fe])/5}
and [hs/ls]\footnote{[hs/ls]=[hs/Fe]-[ls/Fe]} with approximate
errors in the calculation. At the end of the table, we have added
an object  HE ~1305+0007, which was analyzed and found to be a
CEMP r/s star \citep{goswami2006}.
Errors in [ls/Fe], [hs/Fe] and [hs/ls] are calculated using Eq.
\ref{equa}, by taking the respective error for each elements
presented in Tables 1 - 3. For single lines we have used an error
equal to 0.12 dex and for the synthesized abundances 0.2 dex. As
can be noticed from Eq. \ref{equa}, the error increases in each
step, thereby increasing the error in the [hs/ls] index.

\begin {landscape}
\begin{longtable}{lllllllll}
\caption{Elemental abundances: Light elements C, Na, Mg, Ca and Ti}\\ \hline \\
Star & \large[C] &\large[Na~I]   &  \large[Mg~I]   & \large[Ca~I]
 &  \large[Sc~II]  & \large[Ti~I] &  \large[Ti~II]&  Reference\\ \\ \hline
\endfirsthead
\multicolumn{7}{c} {{ \tablename\ \thetable{} -- continued from previous page}} \\ \hline \\
Star &\large[\rm C] &\large[Na~I]   &  \large[Mg~I]   & \large[Ca~I]
 &  \large[Sc~II]  & \large[Ti~I] &  \large[Ti~II] &  Reference \\ \\  \hline
\endhead
\hline \\
\endlastfoot
{\bf Group I}\\
HD~26&7.7 &  5.35  &  6.98&  6.44$\pm$0.11&  3.73$\pm$0.06&  3.96$\pm$0.30&-&1   \\
HD~5223  &  7.9  &   4.57  &  6.05 &  4.35    &   -  &   3.35  &   - &2  \\
$^*$HD~16458     & 8.65 & 6.17$\pm$ 0.14  & 6.92$\pm$ 0.04 & 5.94 & 2.50  & 4.56$\pm$ 0.23 & 4.66$\pm$ 0.28 &3\\
$^\#$HD~122202   & 8.26 &  -    & 7.22 & 6.01 & -     &  -   & 4.63 $\pm$ 0.17&4\\
HD~198269&  8.9 &  4.30&  6.03$\pm$0.09&  4.78$\pm$0.14&  1.51&  3.09$\pm$0.14&  3.33$\pm$0.22&1\\
HD~201626        & 8.22 & 5.19  & 6.83$\pm$ 0.20 & 5.57$\pm$ 0.15 & 1.70  & 4.25$\pm$0.27 & 4.32$\pm$0.25&3 \\
$^\#$HD~204613   & 8.64 & 6.01$\pm$0.24  & 7.40$\pm$0.16  & 6.20$\pm$0.18 & 2.98  &4.92$\pm$0.24 & 5.07$\pm$0.20&4 \\
HD~209621 &  7.7   &  4.10 &  5.76  &  4.45   &  1.92   & -      & 3.70  &5\\
$^\#$HD~216219   & 8.55 & 6.32$\pm$0.12  & 7.51$\pm$0.20 & 6.28$\pm$0.22&2.70  & 4.84$\pm$0.20 & 5.02$\pm$0.27 &3 \\
HD~224959&  8.0&  3.95&  5.50&  4.29$\pm$0.06&-&  2.92$\pm$0.15 &  2.97$\pm$0.21 &5\\

\hline
\bf{Group II}\\

$^\#$HD~4395 & 8.09 &   6.24$\pm$0.19 & 7.49$\pm$0.28& 6.16$\pm$0.22 & 2.80&4.78$\pm$0.22& 4.96$\pm$0.24&3 \\
$^*$HD~55496& 7.91    & 5.08 &  6.37&  5.28&  - &    3.31 & 3.25&4 \\
$^*$HD~48565& 8.39   & 5.76$\pm$0.04  & 7.09$\pm$0.23 & 5.83$\pm$0.22 & 2.35 & 4.35$\pm$0.07 & 4.61$\pm$0.23&3\\
HD~92545    &   8.86   &5.87$\pm$0.03 & 7.23$\pm$0.12& 6.05$\pm$0.23 &-  &   4.72$\pm$0.20 & 5.20$\pm$0.24 &4 \\
HD~104979  &  8.16     &5.90$\pm$0.04& 7.34$\pm$0.06 & 6.10$\pm$0.24 & 2.78& 4.78$\pm$0.16 & 4.96$\pm$0.20&4 \\
HD~107574  &   8.21    &6.01$\pm$0.18 & -    & 5.85$\pm$0.09 & 2.27 &4.57$\pm$0.20 & 4.60&4\\
\hline
\bf{Group III}\\

$^*$HD~5395&   8.09   & 5.54$\pm$0.14 & 7.41$\pm$0.27 & 6.09$\pm$0.18 &2.85& 4.74$\pm$0.21 & 4.84$\pm$0.25&3 \\
HD~81192&  -   &5.38$\pm$0.15 & 7.32$\pm$0.11 & 5.99$\pm$0.19 &2.8& 4.65$\pm$0.20 & 4.65$\pm$0.20&3 \\
HD~89668    &   8.31   &6.12$\pm$0.23 & 7.68 & 6.81$\pm$0.17&  2.92  & 4.51$\pm$0.20& 4.39$\pm$0.22&4\\
HD~111721  &  7.36  &  5.13$\pm$0.14 & 6.88$\pm$0.16 & 5.61$\pm$0.21 & -  &   4.25$\pm$0.24 & 3.82$\pm$0.25 &4\\
HD~125079&8.39    & 6.33$\pm$0.09  & 7.32$\pm$0.27 & 6.16$\pm$0.20 & 2.89&  5.08$\pm$0.25 & 5.17&3\\
HD~126681 &   -    & 5.01$\pm$0.14 &7.07 & 5.48$\pm$0.20 & -  &   4.50$\pm$0.01 & 4.60$\pm$0.15&4\\
HD~148897 &   7.16     &4.86$\pm$0.14&  7.14$\pm$0.07 &  5.48$\pm$0.23 & 1.70 & 3.99$\pm$0.27  & 4.34$\pm$0.20&4  \\
HD~164922&   8.47      &6.38$\pm$0.27  & 8.12 & 6.47$\pm$0.20 &  2.81  & 5.38$\pm$0.19 & 5.18$\pm$0.23&4 \\
HD~167768 & 7.91      & 5.70$\pm$0.25 & 7.19$\pm$0.26 & 6.02$\pm$0.22  & 2.91 & 4.56$\pm$0.22 & 3.90$\pm$0.28 &4\\
HD~188650& 8.09   & 6.47$\pm$0.09  & 7.34$\pm$0.11 & 5.83$\pm$0.20 & 2.05 & 2.83$\pm$0.10 & 2.83$\pm$0.20&3 \\
HD~214714&  8.09 & 6.29$\pm$0.15  & 7.18$\pm$0.21  & 6.13$\pm$0.20 & 2.16& 4.36$\pm$0.20& 4.89$\pm$0.15&3\\
HE~1152-0355& 7.7  &   -    & 6.25  &  -   &    -     &    -  &  4.15  &2 \\
\hline
\\
HE~1305+0007&8.20 &4.40 &   5.75& 4.41& 1.15&-& 3.70 & 2\\
\end{longtable}
{\footnotesize
$^\#$ Sub-giant CH stars\\
$^*$ Objects are also included in Ba star catalogue of Lu 1991\\
References: 1. \cite{goswami2016}, 2. \cite{goswami2006}, 3.
\cite{kari2014}, 4. \cite{kari2015}, 5. \cite{goswami2010}.\\}

\end {landscape}
\begin {landscape}
\begin{longtable}{lcccccccc}
\caption{Elemental abundances: Light elements V, Cr, Mn, Co, Ni and Zn}\\ \hline \\
Star &\large[V~I]  &  \large[Cr~I]  & \large[Cr~II] &
\large[Mn~I] & \large[Co~I]  & \large[Ni~I]
 & \large[Zn~I] &Reference\\ \\ \hline \\
\endfirsthead
\multicolumn{7}{c} {{ \tablename\ \thetable{} -- continued from previous page}} \\ \hline \\
Star &
\large[V~I]  &  \large[Cr~I]  & \large[Cr~II] &
\large[Mn~I] & \large[Co~I]  & \large[Ni~I]
 & \large[Zn~I] &Reference\\ \\  \hline    \\
\endhead
\hline
\endlastfoot
{\bf Group I}\\
HD~26&     -&  4.34&     -&  4.05&    - &  5.06&  3.49&1\\
HD~5223      &     - &     -    &    -   &   -     & 3.70 &   -     &-&2\\
$^*$HD~16458     & 3.60 &  5.28$\pm$ 0.26  & -    & 4.80 & 4.76$\pm$ 0.18 & 5.73$\pm$ 0.25 &4.38$\pm$ 0.20&3\\
$^\#$HD~122202   & -     &  5.12  &-   &  -  &   -   &  5.74 $\pm$ 0.16&  4.56&4\\
HD~198269&2.1&  3.53$\pm$0.18&-&  3.23$\pm$0.16&- &  4.19$\pm$0.14&    -&1 \\
HD~201626         & -     & 3.66$\pm$0.25  & -    &- & 3.53 & 4.73$\pm$0.25& 3.17&3\\
$^\#$HD~204613   &3.75 & 5.34$\pm$0.28  &5.48$\pm$0.25  & 4.82 &4.49$\pm$0.07& 6.03 &  -&4\\
HD~209621    &   -   &    3.50&    -   &     -  &     -  &  4.28 &  2.90&5\\
$^\#$HD~216219    & 3.70 & 5.38$\pm$0.20  & 5.59  &4.90 &4.78$\pm$0.25 & 6.40$\pm$0.23& 4.59$\pm$0.21&3 \\
HD~224959 & -&  2.98$\pm$0.16& -& -& -&  3.90& -&1\\
\hline
\bf{Group II}\\
$^\#$HD~4395 & 3.70& 5.49$\pm$0.29 & 5.63$\pm$0.03  &5.00& 4.57$\pm$0.25& 6.05$\pm$0.21& 4.62&3\\
$^*$HD~55496&2.70& 3.80 &3.94 &-   &  -   &  4.56& 3.13&4\\
$^*$HD~48565& 3.20  & 4.86$\pm$0.15 & 4.73$\pm$0.02  & -0.42& -0.15$\pm$0.25& -0.17$\pm$0.20 & -0.18$\pm$0.13&3\\
HD~92545     &  -   &  5.28$\pm$0.25 & -  &   -   &  5.51$\pm$0.11  & 6.03$\pm$0.12 &-&4\\
HD~104979   & 3.79 & 5.36$\pm$0.23& 5.44$\pm$0.24  &4.90& 4.94$\pm$0.13 & 6.01$\pm$0.24  & 4.31$\pm$0.20&4\\
HD~107574  &     -   &  5.09$\pm$0.25 &  4.73& -   &  -  &   5.59$\pm$0.12   & -&4\\
\hline
\bf{Group III}\\
$^*$HD~5395& 3.52    & 5.21$\pm$0.06 &  5.88  &4.65 & 4.88$\pm$0.20 & 5.90$\pm$0.21& 4.65 &3\\
HD~81192& 3.60  & 4.81$\pm$0.19 & 4.91$\pm$0.19 &4.49&  4.70$\pm$0.08 &5.81$\pm$0.18& 3.98$\pm$0.22&3\\
HD~89668    & -   &  5.42$\pm$0.23 & 5.24 &5.60 & 4.56$\pm$0.25 &5.98$\pm$0.23&-&4\\
HD~111721  & -   &  4.32$\pm$0.20& 4.22$\pm$0.14& -  &   -   &  5.05$\pm$0.22 & -&4\\
HD~125079 & 3.94 & 5.40$\pm$0.25 & -    & 4.99 & 4.77$\pm$0.20 & 6.12$\pm$0.24 &  4.55$\pm$0.11&3\\
HD~126681 &  -  &   4.84$\pm$0.25 &  -   &  -   &  -    & 5.25$\pm$0.23 & -&4\\
HD~148897   & 2.80 & 4.40$\pm$0.20& 4.62  & 3.83& 3.94$\pm$0.15  &5.08$\pm$0.20 & 3.86$\pm$0.18&4\\
HD~164922&    4.63 & 5.91$\pm$0.17& 5.93$\pm$0.03 & 5.76  &5.28$\pm$0.25 & 5.28$\pm$0.25  & 4.98$\pm$0.26&4  \\
HD~167768 & 5.33 &  5.04$\pm$0.20 & 4.92& 4.32& 4.38$\pm$0.11 & 5.63$\pm$0.27 & 4.29$\pm$0.03&4\\
HD~188650& 3.30 & 5.14$\pm$0.23 & 5.19  &4.69 & 4.52$\pm$0.27 & 5.61$\pm$0.17& 4.11$\pm$0.04&3\\
HD~214714 & 3.21  & 4.97$\pm$0.20  & 5.06$\pm$0.15 &4.55& 4.51$\pm$0.25& 5.62$\pm$0.17&  4.14&3\\
HE~1152-0355  &   -   &       -  &     -   &     -  &    -  &     -   &   -&2 \\
\hline
\\
HE~1305+0007   & -    &     -    &   -    &   3.50  &  -    & -3.95   & -&2\\
\end{longtable}
{\footnotesize
$^\#$ Sub-giant CH stars\\
$^*$ Objects are also included in Ba star catalogue of Lu 1991\\
References: 1. Goswami et al. (2016), 2. Goswami et al. (2006), 3. Karinkuzhi and Goswami (2014), 4. Karinkuzhi and Goswami (2015),
5. Goswami \& Aoki (2010).\\
}
\end{landscape}
\begin{landscape}
\begin{small}
\begin{longtable}{lllllllllllll}
\caption{Elemental abundances : Heavy elements}\\ \hline \\
 Star  & [Sr~I] & [Y~II] & [Zr~II] &  [Ba~II] &  [La~II] & [Ce~II] &  [Pr~II] &  [Nd~II]  &  [Sm~II]  &  [Eu~II] & [Dy~II]&Reference\\ \\  \hline  \\
\endfirsthead
\multicolumn{7}{c} {{ \tablename\ \thetable{} -- continued from previous page}} \\ \hline \\
Star  & [Sr~I] & [Y~II] & [Zr~II] &  [Ba~II] &  [La~II] & [Ce~II] &  [Pr~II] &  [Nd~II]  &  [Sm~II]  &  [Eu~II] & [Dy~II]&Reference \\ \\    \hline    \\
\endhead
\hline
\endlastfoot

\bf{Group I}\\
HD~26&3.7&1.95&2.6&2.95&1.51$\pm$0.05&2.08$\pm$0.13&1.20$\pm$0.17&1.75$\pm$0.10&1.74$\pm$0.30&0.1& -&1\\
HD~5223&2.25&0.80& 2.10 & 1.95& 0.85& 1.75& - & 0.95& 0.65& -& - &2\\
$^*$HD~16458 & 3.64 &3.02$\pm$0.25& 3.11$\pm$0.21& 2.70 &1.90& 2.40$\pm$0.21& 1.88$\pm$0.08 &2.35$\pm$0.17& 2.23$\pm$0.18& 0.53 &-&3\\
$^\#$HD~122202 & -  &  3.02& -    & 1.87  &1.40  & 2.57 & 1.34$\pm$0.14 &-  &  2.50 &-  &  -&4\\
HD~198269&1.95&0.38&1.36$\pm$0.06&0.65$\pm$0.14&1.19$\pm$0.15&0.15$\pm$0.09&0.88$\pm$0.11&0.84$\pm$0.13&-&-&1\\
HD~201626& - &- &- &2.90& 1.50& 2.08$\pm$0.21& 1.41 &2.30$\pm$0.22& 1.25$\pm$0.25& -&0.72&3\\
$^\#$HD~204613 & 4.31& 2.94$\pm$0.11& 3.49$\pm$0.04&  2.97 & 2.10 & 2.58$\pm$0.04 & 1.99& 2.23$\pm$0.20& 2.38$\pm$0.17& 0.34 &2.67&4\\
HD~209621&2.0 &0.65 &2.45&1.95&1.62&1.70&0.95&1.40&0.55&-0.05&-&5\\
$^\#$HD~216219 &4.54 & 3.03$\pm$0.22 &3.39$\pm$0.15 &3.13 & 1.99& 2.43& 1.67$\pm$0.15 &2.17$\pm$0.25 &1.74$\pm$0.20 & 0.41&-&3\\
HD~224959&2.0&0.01&1.82$\pm$0.12&1.21$\pm$0.05&1.44$\pm$0.17&0.64$\pm$0.11&1.33$\pm$0.09&1.36$\pm$0.25&0.09&-&1\\
\hline
\bf{Group II}\\
$^\#$HD~4395 &3.81& 2.67$\pm$0.21& 2.98$\pm$0.22& 2.77& 1.97&  1.81$\pm$0.25  &1.05  &2.06$\pm$0.17  & 1.90$\pm$0.24&  -&-&3\\
$^*$HD~48565& 4.06& 2.70$\pm$0.13 & 2.90$\pm$0.04& 3.10 &2.00& 2.41$\pm$0.17 &1.10$\pm$0.02& 2.37$\pm$0.05 &1.60$\pm$0.02 & 0.22&-&3\\
$^*$HD~ 55496& 2.35& 1.65$\pm$0.19& 1.70  &1.33 & -  &   0.30 & 0.27& -   & -   & -   & -&4\\
HD~92545&   -    &0.23$\pm$0.21& -    & 0.91 & 0.95 & 1.6 &  -   & -   & -  &  -  &  -&4\\
HD~104979 & 3.60& 2.61$\pm$0.23& 3.13 & 2.80  &1.93 & 2.33$\pm$0.25  &1.44$\pm$0.20 &2.25$\pm$0.25 &2.05$\pm$0.24 &0.61 &-&4\\
HD~107574 & -   & 2.57$\pm$0.19& -    & 2.48 & 1.51 & 1.52  & -   & - &   -   & - &   -&4\\
\hline\\
\bf{Group III}\\
$^*$HD~5395 &2.94& 2.02$\pm$0.11& - &1.96 & 1.13 & 1.40$\pm$0.17& 1.26$\pm$0.20& 1.95$\pm$0.12&-& 0.62&2.28&3\\
HD~81192 &2.99& 1.80$\pm$0.11 &2.20$\pm$0.12 & 1.79 &  0.90 &0.92$\pm$0.11 &- &2.00$\pm$0.16  & 1.35$\pm$0.25 &-&1.84&3\\
HD~89668 &  3.79& 2.57$\pm$0.30& -  &   1.74 &2.78 & 2.91 & 2.18 &2.65$\pm$0.20& 2.05& 0.71 &-&4\\
HD~111721 & -   & 1.15$\pm$0.21 &-    & 0.97 &0.33&  2.07$\pm$0.09 &   -   & 2.44$\pm$0.14 & -   & -  &  -&4\\
$^\#$HD~125079& 4.33 &3.08$\pm$0.17& - &3.05&    - &2.41$\pm$0.27 & 1.53& 2.46$\pm$0.14& 1.39 &-&-\\
HD~126681&  -   & 1.31$\pm$0.06 &-    & 1.52 & -    & 1.25 & -   & 1.73$\pm$0.10 & 1.16 &-   & -&4\\
HD~148897&  2.24& 1.25$\pm$0.17 &1.13& 0.53& 0.43  &0.43$\pm$0.21& -   & 0.59$\pm$0.21 &0.60$\pm$0.20& -   & 0.17&4\\
HD~164922 &         3.94 & 2.58$\pm$0.30&-&  2.68 &1.51& 1.72$\pm$0.15&- &-&-& -& -&4\\
HD~167768&  3.13& 2.21$\pm$0.24& 2.23$\pm$0.02&   1.25 &0.03 &1.08$\pm$0.30 & -    &1.55$\pm$0.22 &1.35$\pm$0.24& 0.22 &1.62&4\\
HD~188650&- &1.71$\pm$0.16 &-& 1.69 &- &1.08$\pm$0.17& 0.81& 1.37 &0.42$\pm$0.14& -&-&3\\
HD~214714 &-&2.08$\pm$0.20 & 1.96$\pm$0.27 &1.51& -& 1.28$\pm$0.17& 1.29$\pm$0.28 &1.46$\pm$0.24& 1.20$\pm$0.16 &-&-&3\\
HD~1152-0355& -& 1.05& 1.32& 2.45& 1.40& -& -- & 0.58& 0.58&-& -   &2\\
\hline
\\
HE~1305+0007& 1.75& 0.95& 2.65& 2.50& 1.70& 2.12& 1.10& 2.05& 1.62& 0.50& -       &2\\
\end{longtable}
{\footnotesize
\hspace{-0.5cm}$^\#$ Sub-giant CH stars\\
$^*$ Objects are also included in Ba star catalogue of lu91\\
References: 1. Goswami et al. (2016), 2. Goswami et al. (2006), 3. Karinkuzhi and Goswami (2014), 4. Karinkuzhi and Goswami (2015),
5. Goswami \& Aoki (2010).\\
}
\end{small}
\end{landscape}

{\footnotesize
 \begin{table}
\caption{Errors in [ls/Fe] and [hs/Fe]}
\centering
\label{Table:1}
\begin{tabular}{lrrrr}
\hline
Star & [Fe/H]&  [ls/Fe]&  [hs/Fe]& [hs/ls]  \\

\hline\\

\bf{Group I}\\
HD~26&-1.11&1.31$\pm$ 0.34& 1.69$\pm$ 0.44& 0.38 ($\leq\pm$0.50)\\
HD~5223&-2.06&1.19$\pm$ 0.28&1.78 & 0.59 ($\leq\pm$0.39\\
HD~16458 &  -0.65   & 1.34 $\pm$ 0.30    &  1.50$\pm$ 0.45   &  0.16 ($\leq\pm$0.45)         \\
HD~122202& -0.63 & 1.44$\pm$ 0.20 &1.16$\pm$ 0.28 &-0.28($\leq\pm$0.28)\\
HD~198269&-2.03& 0.64$\pm$ 0.28&1.55$\pm$ 0.33& 0.91 ($\leq\pm$0.44)\\
HD~201626& -1.39  &  -      & 1.93$\pm$ 0.45    &  -         \\
HD~204613& -0.24 & 1.27$\pm$ 0.23 &1.16$\pm$ 0.42& -0.11($\leq\pm$0.42)\\
HD~209621&-1.94   & 0.79$\pm$ 0.28 & 1.89$\pm$ 0.28 & 1.1 ($\leq\pm$0.45)         \\
HD~216219& -0.17  & 1.26$\pm$ 0.33    & 1.01$\pm$ 0.34    & -0.25($\leq\pm$0.34)     \\
HD~224959&-2.42& 0.86$\pm$ 0.28 & 2.24$\pm$ 0.34&1.38 ($\leq\pm$0.45)\\
\hline
\bf{Group II}\\
HD~4395  & -0.18  & 0.77$\pm$ 0.20    &  0.82$\pm$ 0.20   &  0.05($\leq\pm$0.20)      \\
HD~48565 & -0.59  & 1.24$\pm$0.31    & 1.47$\pm$0.42    & 0.23 ($\leq\pm$0.42)     \\
HD~55496  &-1.49 & 0.73$\pm$0.27 &0.38$\pm$ 0.41&-0.35($\leq\pm$0.41)\\
HD~92545  &-0.21 & 0.23$\pm$0.29 &1.15$\pm$0.20& 0.92($\leq\pm$0.29)\\
HD~104979& -0.26 & 0.85$\pm$0.30 &1.03$\pm$0.51& 0.18($\leq\pm$0.51)\\
HD~107574& -0.48 & 1.02$\pm$0.27 &0.87$\pm$0.29& -0.15($\leq\pm$0.29)\\
\hline
\bf{Group III}\\
HD~5395  & -0.24  & 0.16$\pm$0.22    & 0.27$\pm$0.44    & 0.11($\leq\pm$0.44 )      \\
HD~81192 & -0.50  & 0.26$\pm$0.21    & 0.34$\pm$0.27    & 0.08 ($\leq\pm$0.27)    \\
HD~89668  &-0.13 & 0.81$\pm$0.48 &1.16$\pm$0.32 &0.35($\leq\pm$0.48)\\
HD~111721& -1.11 & 0.05$\pm$0.29 &0.98$\pm$0.20 &0.93($\leq\pm$0.29)\\
HD~125079&  -0.18 &  1.32$\pm$0.23  & 0.93$\pm$0.25   &  -0.39($\leq\pm$0.25)     \\
HD~126681& -0.90 & 0.02$\pm$0.20 &0.80$\pm$0.34 &0.78($\leq\pm$0.34)\\
HD~148897& -1.02 &-0.13$\pm$0.28 &0.01$\pm$0.39& 0.14($\leq\pm$0.39)\\
HD~164922&0.22 & 0.47$\pm$0.36 & 0.10$\pm$0.38 & -0.37($\leq\pm$0.38)\\
HD~167768& -0.51 & 0.51$\pm$0.31 &0.14$\pm$0.42& -0.37($\leq\pm$0.42)\\
HD~188650 & -0.45  & -0.03$\pm$0.20& 0.23$\pm$0.20    & 0.26($\leq\pm$0.20)        \\
HD~214714& -0.35  & -0.03$\pm$0.21 & 0.14 $\pm$0.43   &  0.17($\leq\pm$0.43)     \\
HE~1152-0355&-1.27& 0.07$\pm$0.28 & 1.11$\pm$0.28& 1.04($\leq\pm$0.39)\\
\hline
\\
HE~1305+0007&-2.0& 1.11$\pm$0.28 & 2.52$\pm$0.28 & 1.30 ($\leq\pm$0.39) \\
\hline
\end{tabular}

The errors given in the 5th column in parentheses are the upper limits of errors in [ls/Fe] and [hs/Fe].\\

\end{table}
}

\section{Theoretical models} \label{modtheo}

Theoretical models presented in this paper have been computed with
the FUNS (FUll Network Stellar) evolutionary code \cite[][and
reference therein]{stra06}. In the last 15 years, main upgrades to
the code have been:
 \begin{enumerate}
 {\item {\it Treatment of convective borders}: in FUNS, convective borders
are fixed by the Schwarzschild criterion, while the temperature
gradient in the convective unstable regions is computed in the
framework of the Mixing Length Theory in the formulation by
\citet{coxgiuli}. When the H-rich opaque convective envelope
penetrates the underlying He-rich radiative region (i.e. during a
TDU), a discontinuity arises at the convective/radiative
interface. If we perturb such a border, we would find it unstable.
In order to overcome this problem, we hypothesize that convective
eddies penetrate beyond the formal Schwarzschild border and we
mimic it by adding an exponentially decaying profile of velocities
at the base of the convective envelope (see \citealt{cri09,cri11}
for its calibration; see also \S \ref{refere}). Such an algorithm
smooths the radiative temperature gradient, thus removing the
aforementioned discontinuity.} {\item {\it $^{13}$C pocket
formation}: the introduction of the exponentially decaying profile
of convective velocities implies two further consequences: a more
efficient TDU and the mixing of a limited amount of protons in the
underlying \cd-rich layers. We limit the extension of this
extra-mixed zone to two pressure scale heights (2 H$_P$), starting
from the formal Schwarzschild border. When this region heats up, a
tiny \ct enriched layer forms. The upper region of the \ct pocket
is partially overlapping with a \nq pocket, which acts as a strong
neutron poison via the \nqnp reaction. The larger the mass of the
H-exhausted core, the thinner the \ct pocket. Thus, in low mass
models the pockets formed after the first TPs dominate the
following \s-process nucleosynthesis. As the initial stellar mass
increases, the predominance of the \ctan neutron source is
progressively substituted by \nean. In fact, the core mass growth
implies a reduction of the \ct pocket mass extension and an
increase of the maximum temperature attained at the base of the
convective shells during TPs. Therefore, the resulting surface
s-process distribution strongly depends on the initial stellar
mass (see \citealt{cri2015b}).}
 {\item {\it Mass-loss formula}: For the pre-AGB phase we use the classical
Reimers' mass-loss rate (by setting $\eta$ = 0.4), while for the
AGB phase we calibrate the mass loss rate on the period - K band
luminosity relation proposed by \cite{white03}, who studied a
sample of O-rich and C-rich giant stars in the Magellanic Clouds
(see \citealt{stra06} for details). We use the same prescription
for the whole range of initial stellar masses and metallicities.}
 {\item {\it C-rich molecular opacities}: depending
on the C/O ratio, O-bearing or C-bearing molecules represent the
dominant opacity source in the coolest layers of the convective
envelope (see e.g. \citealt{marigo2002}). Since the carbon
reservoir of the envelope increases with the TDU number, the
opacity of those layers increases as well, implying more expanded
and cooler structures \citep{cri07}. As a consequence, the use of
C-rich molecular opacities leads to a larger mass loss rate with
respect to a case in which a solar-scaled distribution is used to
calculate opacities.}
 {\item {\it Rotation}: the effects
induced by rotation on the \s-process nucleosynthesis are
discussed in more detail in \cite{pi13}. Here we only recall that
the combined effect of Eddington-Sweet circulations and
Goldreich-Schubert-Fricke instabilities smears off the profiles of
the \ct and \nq pockets. As a consequence, the {\it
neutron-to-seed} ratio is lower and, thus, the efficiency of the
\s-process decreases. Our reference models set is computed without
rotation. In \S \ref{rotate} we present a new set of 1.5 M$_\odot$
rotating AGB models at different metallicities.}
 {\item {\it Nuclear network}: the physical evolution of FUNS models
is coupled with a full nuclear network (from hydrogen to lead, at
the termination of the \s-process path). This allows us to
directly follow the s-process nucleosynthesis avoiding
post-process calculations. Our network includes about 500 isotopes
linked by more than 1000 reactions \citep{cristallophd,stra06}.}
\end{enumerate}

Our theoretical models (those previously published and those
presented in this paper) are available as a part of the online
FRUITY (FUNS Repository of Updated Isotopic Tables \& Yields)
database\footnote{http://www.oa-teramo.inaf.it/fruity}
\citep{cri11,cri2015b}. From this web platform it is possible to
freely download the chemical (elemental and isotopic surface
distributions, yields, s-process indexes, etc.) as well as the
physical features of the models (masses, luminosities, gravities,
number of Thermal Pulses and TDUs, TDU efficiency, etc.).

\section{Comparison between observations and FRUITY models} \label{refere}

\begin{figure*}[tpb]
\centering
\includegraphics[width=\textwidth]{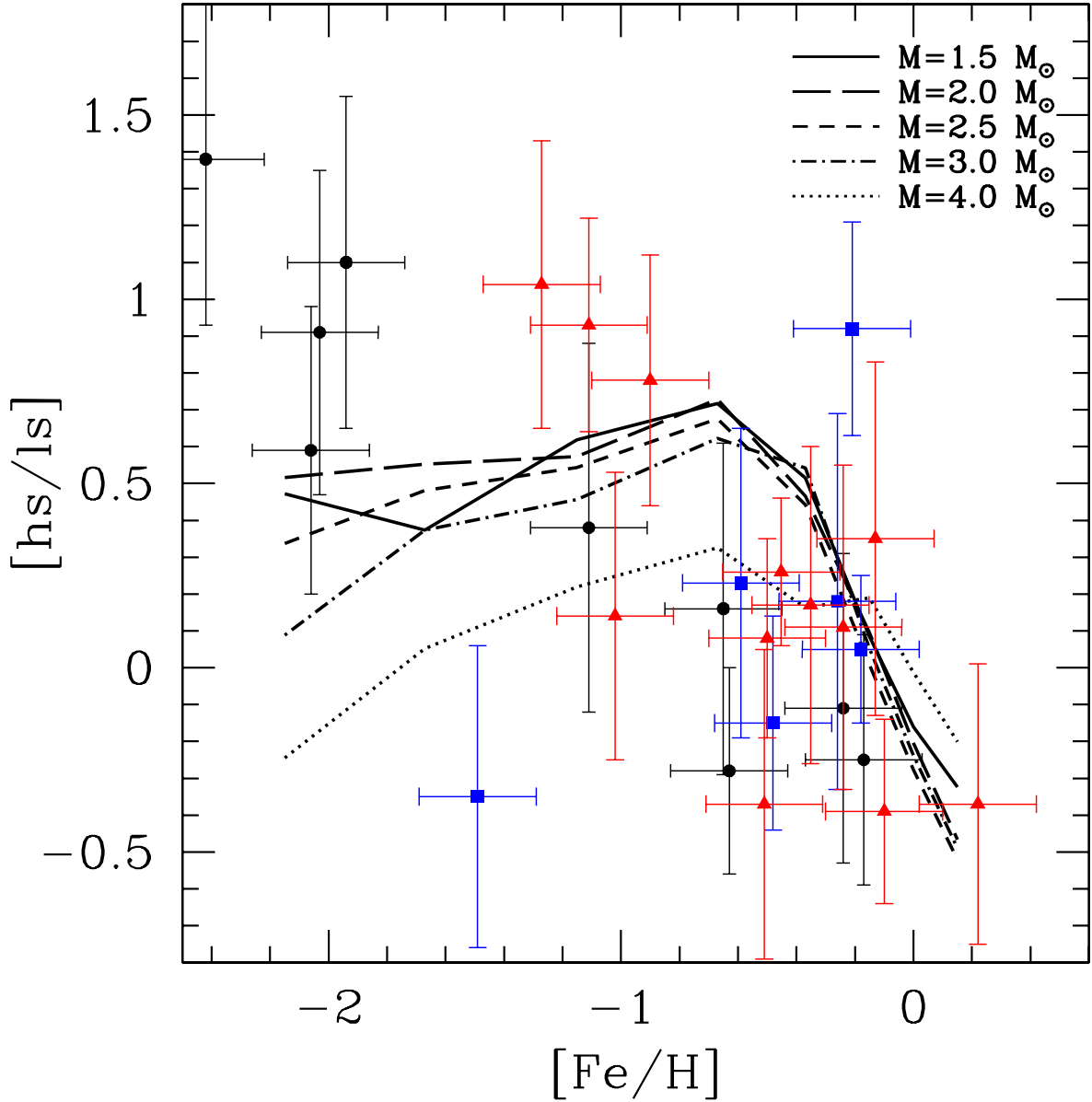}
\caption{[hs/ls] ratios of our CH sample as a function of the
metallicity compared to FRUITY models. Symbols refer to the three
groups identified in Table 4: circles (Group I), squares (Group
II) and triangles (Group III).  } \label{fig1}
\end{figure*}

\begin{figure*}[tpb]
\centering
\includegraphics[width=\textwidth]{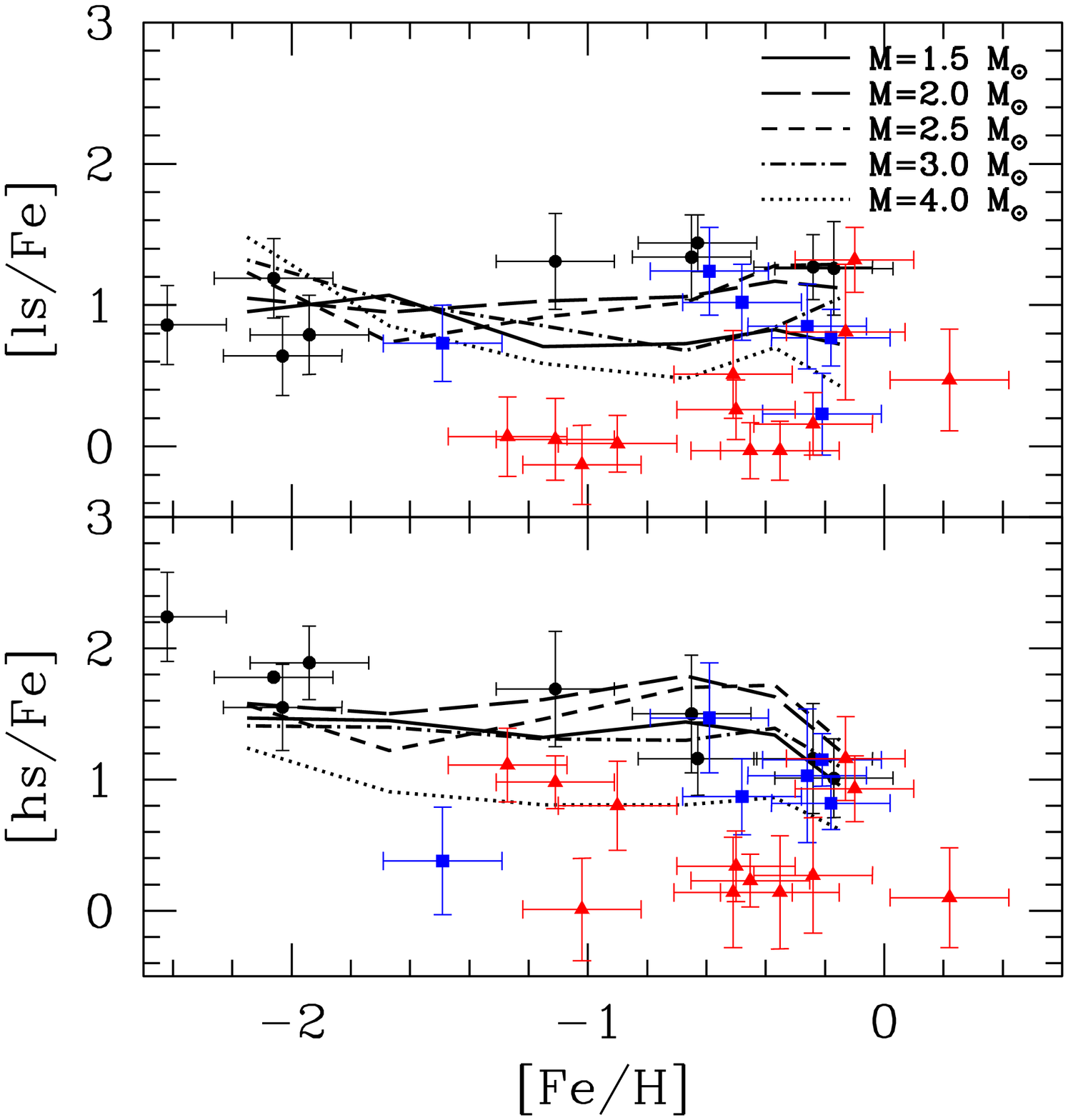}
\caption{As in Figure \ref{fig1}, but for [ls/Fe] and [hs/Fe]
(upper and lower panel, respectively).} \label{fig2}
\end{figure*}

In Figure \ref{fig1} we report the [hs/ls] ratios of our CH sample
(see Table 4) as a function of the iron content [Fe/H]. We plot
the three groups with different symbols: circles (Group I),
squares (Group II) and triangles (Group III). According to this
plot, the groups seem to represent a homogeneous sample. We focus
on the [hs/ls] s-process index because our low mass models attain
a nearly asymptotic value after a few TDUs. Therefore, our [hs/ls]
ratios are almost independent from the adopted mass-loss rate (see
the discussion in \citealt{cri11}), thus reducing the degeneracy
in the free parameter space. The theoretical behavior of the
models is clear: starting from the highest computed metallicity,
there is an increase of the [hs/ls] value down to
[Fe/H]$\sim$-0.7. This is due to the fact that, at large
metallicities, mostly ls elements are synthesized, while with
decreasing Z the {\it neutron-to-seed} ratio increases and the hs
component is produced too. At lower [Fe/H], lead starts being
produced, basically freezing the [hs/ls] ratio to a value of about
0.5-0.6. Apart from the 4 M\odo star, in which the \nean
contribution dominates, the other models present quite similar
[hs/ls] ratios for a fixed metallicity. Observations show a
different trend. Apart from a couple of isolated stars, two
distinct features appear: an increasing [hs/ls] ratio down to very
low metallicities and a definitely larger spread for a fixed
[Fe/H].
A further criticality may arise when comparing the theoretical
[ls/Fe] and [hs/Fe] to observations (see Fig. \ref{fig2}). As
already stressed before, CH stars owe their s-process surface
enrichment to a transfer episode from the already extinct AGB
companion. Many of our CH stars are in fact confirmed binaries
(those belonging to Group I) and others have known radial velocity
variations (those belonging to Group II). Furthermore, when
comparing theoretical models to observation, a possible dilution
due to the evolutionary status of the accreted star may be taken
into account (see e.g. \citealt{bi11}), as well as long-lasting
processes, such as gravitational settling (see e.g.
\citealt{stancliffe2010}). It is far from the purpose of this
paper to determine the binary system properties of our CH sample.
However, we can provide some general considerations based on first
principles. The degree of the dilution depends on the evolutionary
phase of the observed CH star. Depending on the initial mass,
these objects may have no convective envelope during the Main
Sequence phase: in this case, no dilution has to be explicitly
taken into account. The situation is different if the envelope is
penetrating downward while ascending the Red-Giant Branch
(Sub-Giant stars) or if it is already receding (Red-Giant stars).
In the second case, FDU has already mixed the whole envelope and
the dilution factor definitely grows (up to 1 dex). FRUITY models
with initial mass M$\sim$(2.0-2.5) M\odo show large enough final
surface s-process enhancements to account for the majority of the
observed stars. However, our models may underestimate the
s-process overabundances in case of a large dilution. It is worth
noticing that, according to Fig. \ref{fig2}, a non negligible
fraction of CH stars belonging to Group III (triangles) shows no
absolute surface s-process enhancement (within the errors).
Therefore, some of these objects may not be CH stars, although
they are included in the catalog of \cite{bartkevicius1996}.

\section{The effects of rotation} \label{rotate}

In \S \ref{modtheo} we briefly mentioned the effects that rotation
has on the s-process nucleosynthesis. \cite{pi13} demonstrated
that the mixing induced by rotation (mainly Edddington-Sweet
circulations and Goldreich-Schubert-Fricke instabilities) leads to
a decreasing {\it neutron-to-seed} ratio with increasing initial
rotation velocity (v$_{ini}^{rot}$). For a fixed initial mass and
metallicity, the larger v$_{ini}^{rot}$, the lower the [hs/ls] and
the [Pb/hs] ratios. This is evident in Figure \ref{fig4}, where we
compare the reference 1.5 M\odo models FRUITY set (labelled REF)
to an analogous set of rotating models with v$_{ini}^{rot}$=60
km/s (labelled V60). Major effects occur at [Fe/H]$\sim$-0.8,
showing an [hs/ls] difference larger than a factor of 10. At low
metallicities, such a difference is less evident because, in this
case, lead is the most affected s-process element (see Figure 9 of
\citealt{pi13}). Unfortunately, lead could not be detected in the
majority of the spectra in our sample, otherwise it would be the
ideal indicator to verify the presence of rotation in the donor
AGB stars.

Depending on the environmental formation process, a star could
approach the Main Sequence with different initial rotation
velocities. In principle, the larger the mass is, the larger the
expected rotation velocity is. Moreover, the lower the metallicity
is, the higher the initial rotation velocity is (because objects
are more compact). As a consequence, the effect of rotation of the
ongoing nucleosynthesis may differ from star to star. It is
therefore reasonable to hypothesize that a spread in
v$_{ini}^{rot}$ leads to a spread in the s-process efficiency.
However, independently from v$_{ini}^{rot}$, the [ls/Fe] and
[hs/Fe] indexes of rotating models are lower than the
corresponding non rotating cases (see lower panel of Fig.
\ref{fig4}). This implies that a large fraction of CH stars cannot
be fitted by our rotating AGB models, even taking into account the
possibility that a larger AGB mass polluted the observed accreted
CH star. In summary, the introduction of rotation alone improves
the fit to the observed [hs/ls] ratios, but absolute surface
s-process values pose severe limitations to the initial rotational
velocity. It is worth to remember that recent asteroseismologic
measurements (see e.g. \citealt{mosser2012}) suggest that the
cores of Red Clump stars (thus in an evolutionary phase prior to
the AGB) have a lower rotation velocity than what we find in our
models. Therefore, the inclusion of some form of angular momentum
re-distribution \citep{marques13} or braking \citep{spruit2002} is
necessary. Among the possible scenarios to slow down stellar
cores, magnetic fields are the most promising candidates. Note, we
also call to mind that low metallicity stars are more compact than
their solar-metallicity counterparts and, therefore, they are
expected to rotate faster for a fixed initial angular momentum.
These considerations have to be taken into account when assigning
the initial v$_{ini}^{rot}$ to a model. Notwithstanding, our goal
is to highlight general trends rather than giving a quantitative
picture of the problem. For this reason we limit our analysis to
just one mass and one rotation velocity.
\begin{figure*}[tpb]
\centering
\includegraphics[width=\textwidth]{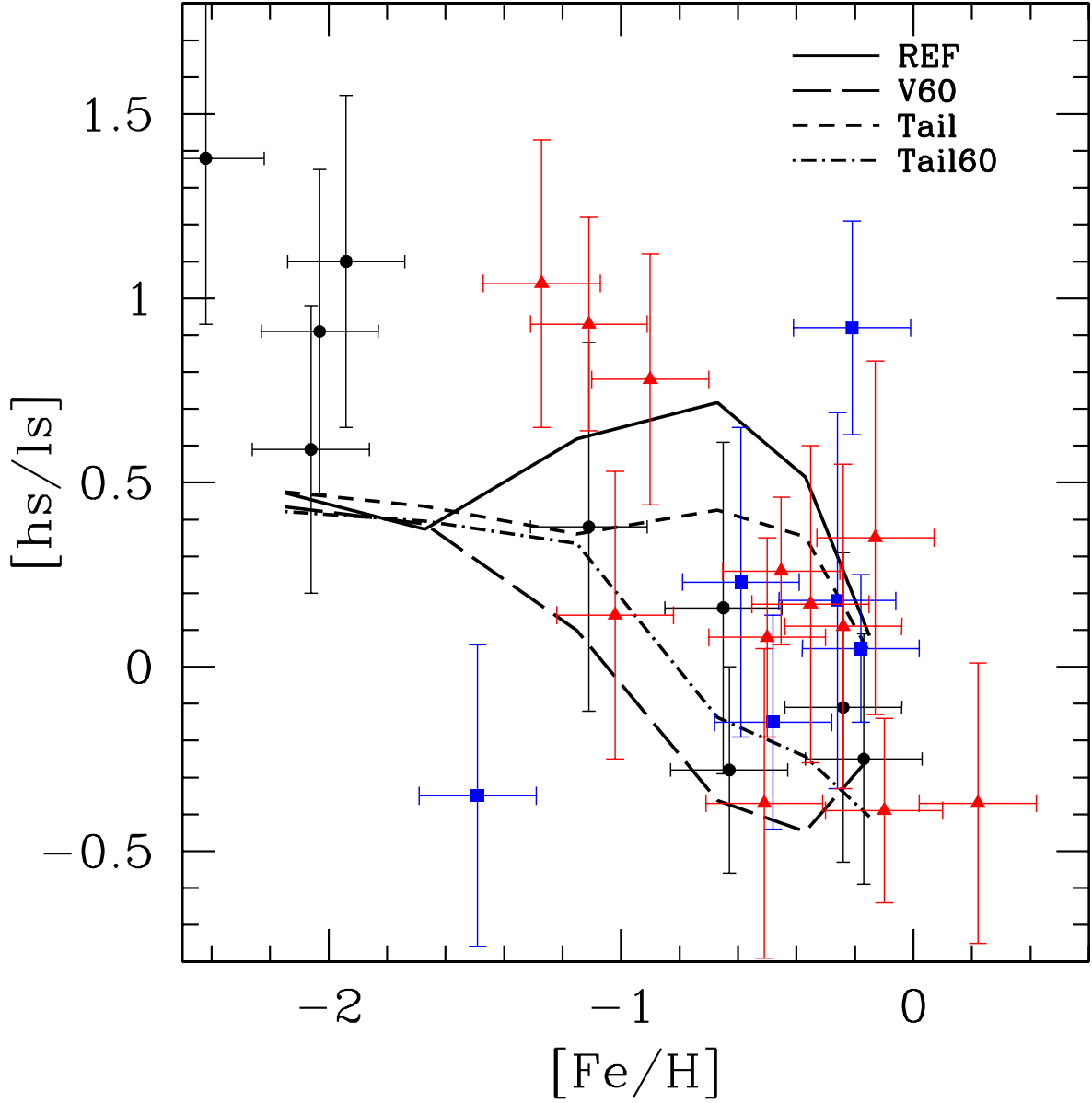}
\caption{[hs/ls] ratios as a function of the metallicity for a set
of 1.5 M\odo stars and different prescriptions for the physics.
Various curves refer to: FRUITY models (REF), rotating models with
v$_{ini}^{rot}$=60 Km/s (V60), non-rotating models with extended
\ct pockets (Tail), rotating models with v$_{ini}^{rot}$=60 Km/s
and extended \ct pockets (Tail60). Observations have the same
symbols as in Fig. \ref{fig1}.} \label{fig3}
\end{figure*}
\section{The effects of a different treatment of the radiative/convective interfaces in AGBs} \label{convetta}

\begin{figure*}[tpb]
\centering
\includegraphics[width=\textwidth]{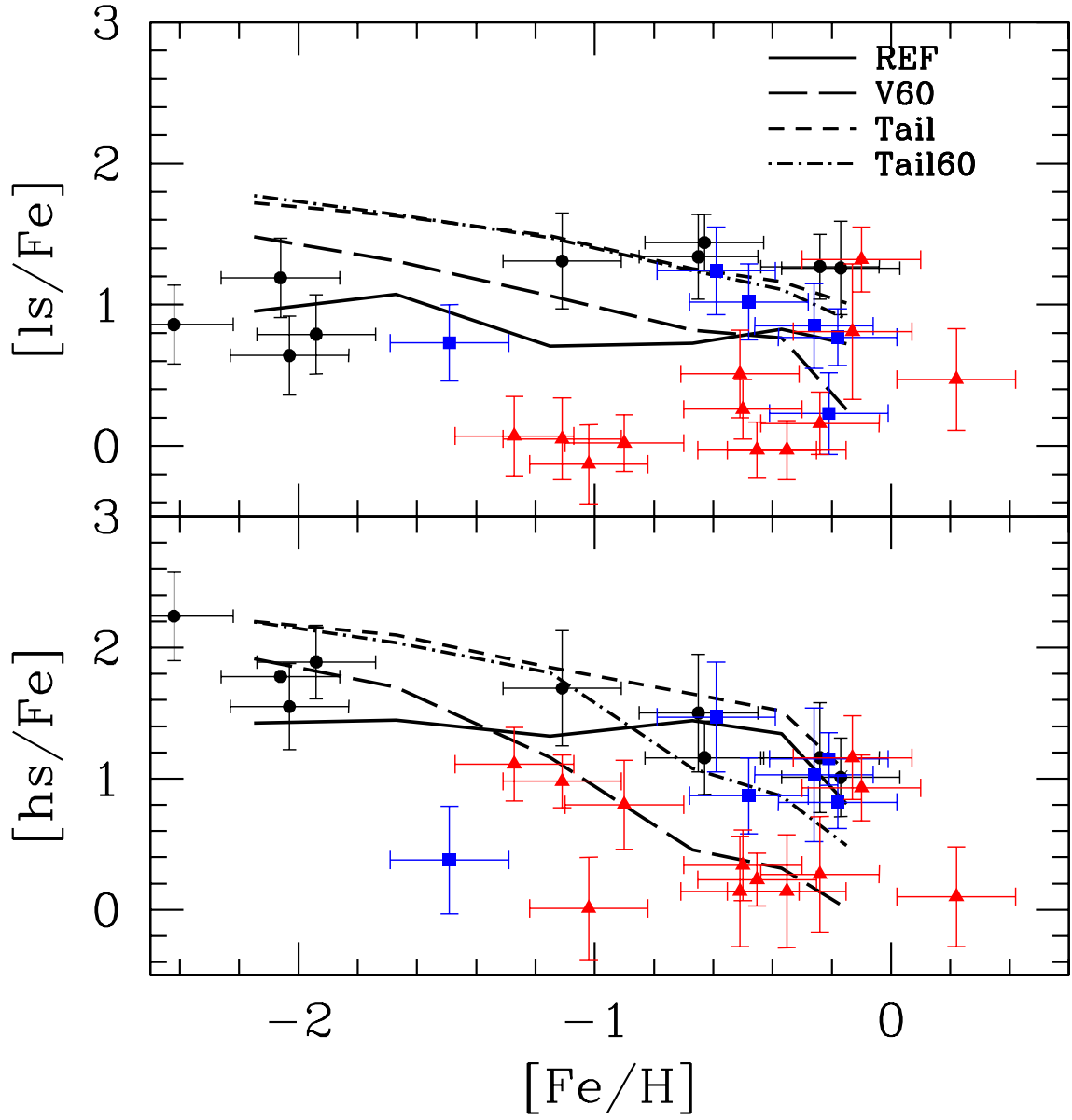}
\caption{As in Figure \ref{fig3}, but for [ls/Fe] and [hs/Fe]
(upper and lower panel, respectively).} \label{fig4}
\end{figure*}

In \S \ref{modtheo} we already explained in detail our
prescription for the handling of the convective/radiative
interface at the inner border of the convective envelope. In
\cite{cri11} we discussed the effects of changing the free
parameter characterizing our exponential velocity profile and we
justified our choice. Recently, we also studied the effects of
relaxing the 2 H$_P$ condition we impose at the lower boundary of
the extra-mixed zone \citep{cri2015a}. In that paper we let the
velocity profile extend down to a layer characterized by a mixing
velocity equal to 10$^{-11}$ times the convective velocity at the
Schwarzschild border (models labelled "Tail" in Fig. \ref{fig3}
and Fig. \ref{fig4}). As a net effect, the \ct pockets forming
after each TDU present a more extended tail with low \ct
abundances and negligible \nq abundances. The need for this kind
of extended \ct pockets has been suggested by the comparison of
theoretical models with open cluster observations
\citep{orazia2009,maiorca2011,maiorca2012} and, more recently,
laboratory measurements of isotopic ratios in pre-solar SiC grains
\citep{liu13,liu14,liu15}. The formation of \ct pockets with
extended tails leads, in our models, to increased surface
s-process overabundances. The differences to the reference FRUITY
set slightly increase with decreasing initial iron content (see
Fig. \ref{fig4}; see also Figure 7 of \citealt{cri2015a}). The
variation in the [hs/ls] ratio is limited, reaching a maximum
difference of about 0.3 dex around [Fe/H]$\sim$-0.8. In summary,
the {\it Tail} models show increased surface final s-process
overabundances without largely modifying the [hs/ls] index.
\subsection{Extended undershooting and rotation}

In Figures \ref{fig3} and \ref{fig4} we present the 1.5 M\odo set
computed with extended \ct pockets and initial rotation velocity
v$_{ini}^{rot}$=60 Km/s (labelled as {\it Tail60}). The [hs/ls]
ratios of the {\it Tail60} set are slightly larger than those
obtained in the {\it V60} set. Therefore, the possibility to
reproduce the observed spread by assuming different rotation
velocities is preserved. Notwithstanding, the disagreement between
theory and observations still exists at [Fe/H]$<$-1. The [ls/Fe]
indexes are very similar to the {\it Tail} case, while the [hs/Fe]
distribution shows a decrease at large metallicities. This effects
could be compensated by an increase of the AGB donor mass (up to
2.5-3.0 M$_\odot$), which would provide a large enough s-process
pollution to reproduce the most enriched CH stars. In summary, the
{\it Tail60} set seem to provide a satisfactory solution for both
the absolute and relative s-process abundances, at least for
[Fe/H]$>$-1. At low metallicities, instead, it clearly emerges
that the hs production obtained in our models is not large enough
to fit observations.

\section{The CEMP case} \label{lowz}

CH stars discussed in the previous Sections belong to a larger
class of objects, i.e. stars enriched by mass-transfer from an
extinct AGB donor. Historically, these objects have been
classified based on their spectrum, quite often coinciding with a
metallicity segregation. At large metallicities, Ba-stars are
observed, while the equivalents of CH stars at low metallicity are
CEMP stars.


CEMP stars represent a significant fraction of the low end of the
Galactic halo metallicity distribution. \cite{bc2005} have
classified these objects into four major groups based on their
heavy element abundance patterns.  The most numerous is the CEMP-s
group, that exhibits enhancement of slow neutron-capture elements
in their surface chemical composition. CEMP stars that are
characterized by the enhancement of r-process elements are called
CEMP-r stars, although this class has very few confirmed examples
so far. Another class of CEMP stars exhibits large overabundances
of elements produced by both s-process and r-process (the
so-called CEMP-rs stars). CEMP stars that do  not show enhancement
of heavy elements form another class (called CEMP-no stars). We
note that, recently \cite{jorissen2016b} demonstrated that CH and
CEMP-s stars lie in the same region of the period - eccentricity
diagram and present similar mass-function distributions.
Therefore, they suggested that these two classes of objects should
be treated as a unique stellar family, as we did in the combined
analysis presented in this paper.

\begin{figure*}[tpb]
\centering
\includegraphics[width=\textwidth]{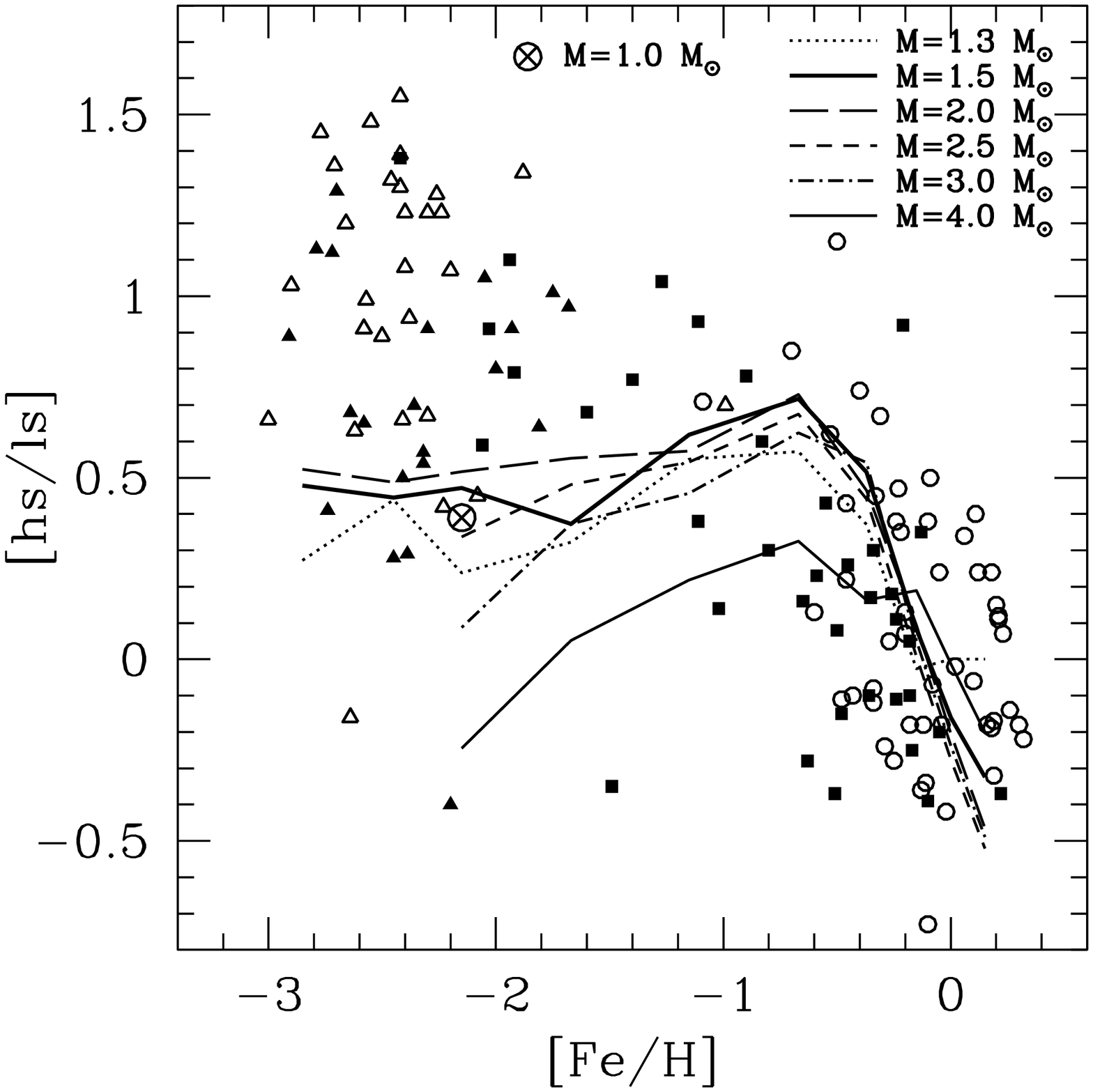}
\caption{Theoretical [hs/ls] as a function of the metallicity for
different masses. Theoretical curves refer to the non rotating
reference FRUITY models. Various symbols refer to: \textbf{Ba
stars} (\citealt{ab2006}, \citealt{smilia2007},
\citealt{liu2009ba}, \citealt{pereira2011},
\citealt{lebzelter2013}; \emph{open circles}); \textbf{CH stars}
(\citealt{smith1993}, \citealt{zacs2000}, \citealt{goswami2006},
\citealt{pereira2009}, \citealt{goswami2010},
\citealt{pereiradrake2011}, \citealt{pereira2012},
\citealt{liu2012ba}, \citealt{kari2014}, \citealt{kari2015},
\citealt{goswami2016}; \emph{filled squares}); \textbf{CEMP-s
stars} (\citealt{preston2001}, \citealt{aoki2002_1},
\citealt{aoki2002_2}, \citealt{lucatello2004},
\citealt{cohen2006}, \citealt{cohen2013}, \citealt{placco2013},
\citealt{roederer2014}; \emph{filled triangles}); \textbf{CEMP-rs
stars} (\citealt{aoki2002_3}, \citealt{vaneck2003},
\citealt{johnson2004}, \citealt{lucatello2004},
\citealt{barbuy2005}, \citealt{ivans2005}, \citealt{cohen2006},
\citealt{jonsell2006}, \citealt{roederer2008},
\citealt{thompson2008} , \citealt{behara2010},
\citealt{masseron2010}, \citealt{roederer2010},
\citealt{cohen2013}, \citealt{cui2013}, \citealt{placco2013},
\citealt{roederer2014}, \citealt{hollek2015},
\citealt{jorissen2016}; \emph{open triangles}).} \label{fig5}
\end{figure*}

\begin{figure*}[tpb]
\centering
\includegraphics[width=\textwidth]{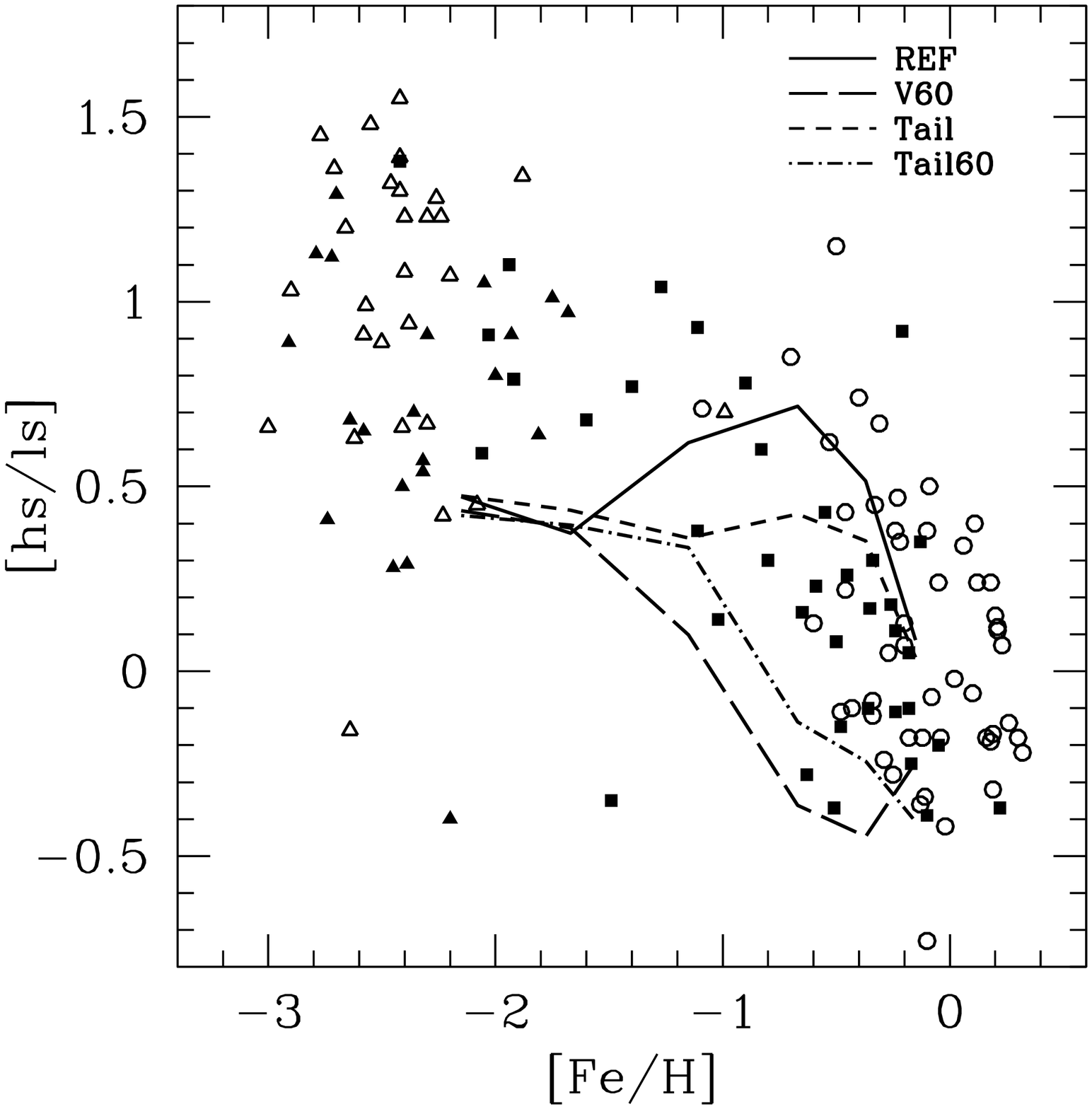}
\caption{As Fig. \ref{fig5}, but for different physics
prescriptions.} \label{fig6}
\end{figure*}

It should be noted that not all CEMP-s stars are binaries. In
fact, \cite{hansen2016} have recently shown from long term radial
velocity monitoring that the binary frequency of CEMP-s stars is
about 82 $\pm$ 10\%. Out of 22 objects that they have studied
eighteen of them exhibit clear orbital motion, while four stars
appear to be single. Thus they confirmed that the binary frequency
of CEMP-s stars is much higher than for normal metal-poor giants,
but not 100\% as previously claimed. We exclude from our analysis
CEMP stars without a clear detection of europium (an r-process
element) in order to
avoid a wrong identification as CEMP-s or CEMP-rs stars.\\
In Fig. \ref{fig5} we report our theoretical [hs/ls] ratios for
different initial masses and compare them to observations (see the
Caption for the references). In order to span over a larger
metallicity grid, we computed additional low mass models
(1.3-1.5-2.0) with [Fe/H]=-2.45 and [Fe/H]=-2.85 (both with
[$\alpha$/Fe]=0.5).
Our set does not reproduce the high [hs/ls] ratios detected in low
metallicity CH and CEMP-s stars ([hs/ls]$\sim$1). The same
conclusion still holds when adopting different prescriptions for
convection and/or rotation (see Fig. \ref{fig6}). Note that an
increase of the initial stellar mass would not improve the
situation. In fact, models with M$\geq$ 2.5 M$_\odot$ at low
metallicities have quite large core masses at the beginning of the
AGB phase and therefore show chemical signatures typical of
intermediate mass models at large metallicities (see, e.g.,
\citealt{cri11}). Notwithstanding, we compute additional 2
M$_\odot$ models with [Fe/H]=-2.45 with rotation and/or extended
tail. With respect to the 1.5 M$_\odot$ models, we find increased
s-process surface enhancements, but almost the same [hs/ls].
Alternative solutions need thus to be found, as for example the
introduction of mixing induced by magnetic fields (see
\citealt{trippa2} and references therein). We will test such a
possibility in a forthcoming paper.

Useful hints on the needed \ct pocket shape can be derived from
theoretical post-processing calculations with constant {\it ad
hoc} \ct pockets \citep{bi10,bi11,bi12}. Those works demonstrated
that at low metallicities [hs/ls]$\sim$1 can be obtained and that
a good fit to a large number of s-process enriched CEMP stars can
be found. In those calculations the \ct pockets are artificially
added after each TDU (with a constant mass extension) and the \ct
abundance is freely varied within the pockets \citep{ga98}. Within
that framework, fits to observations are obtained with \ct pockets
characterized by very low \ct abundances (see Figure 17 of
\citealt{bi11}).

As already mentioned before, there is a class of CEMP stars
showing surface enrichment of elements ascribed to both the
s-process and the r-process (the so-called CEMP-rs stars). Those
objects show the largest observed [hs/ls] ratios (see asterisks in
Fig. \ref{fig5}). Actually, there is no consensus on their
pollution history (see \citealt{abate2016} and references
therein). Among the proposed theories, there is the so-called {\it
i}-process \citep{cowan1977}, i.e. a nucleosynthetic process
providing neutron densities intermediate between those
characterizing the s-process ($n_n \sim$ 10$^{7}$ cm$^{-3}$) and
the r-process ($n_n
>$ 10$^{21}$ cm$^{-3}$). This range of neutron densities can be
attained in low mass low metallicity stars, when protons from the
envelope are ingested into the underlying convective He
intershell. We refer to this episode as the Proton Ingestion
Episode (PIE). The occurrence of such a mixing phenomenon depends
on the initial stellar mass and metallicity: the lower the mass
and the metallicity, the higher the probability for a PIE to
occur. At extremely low metallicities ([Fe/H]$<$-3), a PIE can be
found during the off-center He-burning flash, while for larger Z
(-3$<$[Fe/H]$<$-2) it may occur at the first fully developed TP
\citep{hollowell1990,fujimoto2000,iwamoto2004,cala2008,lau2009}.
3D hydrodynamical simulations confirm this peculiarity, which
characterizes very metal-poor models \citep{woodward2008}. In
2009, we presented the evolution and nucleosynthesis of a 1.5
M\odo model with [Fe/H]=-2.45 and no $\alpha$ enrichment
\citep{cri09b}. That model experiences a strong PIE, with
important consequences for the physical and chemical evolution of
the structure. In particular, we found that soon after the
occurrence of a PIE, a very deep TDU occurs, which carries to the
surface the freshly synthesized \ct, \nq and light s-process
elements. Those nuclei are the result of a convective
high-temperature H-burning. Since the CNO burning occurs in
non-equilibrium conditions, a huge amount of \ct is produced. This
leads to a very efficient neutron production from the \ctan
reaction, with neutron densities reaching $n_n \sim$ 10$^{15}$
cm$^{-3}$. However, the local energy release from in-flight proton
burning leads to the splitting of the convective He-shell and
stops any further growth of the neutron density. It is important
to note that the mass and temporal coordinates of such a splitting
strongly depend on the adopted nuclear network. In fact, the split
occurs when the energy released by proton capture reactions
overwhelms the energy production at the base of the convective
shell (see Figure 3 of \citealt{cri09b}). The energy budget
triggering the convective shell comes from 3$\alpha$ reactions,
whose Q value is about 7 MeV. However, when enough \ct has been
mixed within the shell, the \ctan reaction (Q$\sim$ 2 MeV) and the
following neutron capture (on average Q$\sim$5 MeV) provide an
additional important energetic contribution. This comes from the
fact that, even if the \ct abundance is definitely lower, its
$\alpha$ capture cross section is at least 7 orders of magnitude
larger than that characterizing the 3$\alpha$ process. Therefore,
in order to properly calculate the energetics of PIEs, it is
mandatory to use a complete neutron capture network coupled to the
physical evolution of the stellar structure. In the 2009 model, we
found a very low [hs/ls] ratio after the deep TDU following a PIE.
Later, the following standard TDU episodes re-establish the trend
expected at low Z, i.e. with the hs component exceeding the ls
one. Thus, depending on the initial mass (and, consequently, on
the TDU number), the PIE signature can be masked by the following
standard s-process nucleosynthesis. A way to trace it, however,
exists and it is connected to barium isotopic ratios. The
occurrence PIEs, in fact, pushes the nucleosynthesis path far from
the $\beta$ stability valley. Very neutron-rich isotopes are
produced, such as $^{135}$I (an isotope having a magic number of
neutrons, N=82). This radioactive isotope decays to $^{135}$Cs
and, later, to its stable isobar $^{135}$Ba. Therefore, the
occurrence of a PIE could be proven by the detection of a very
large $^{135}$Ba abundance. Note that the cross section of
$^{135}$I is very low compared to other magic nuclei. Moreover, it
has been determined only theoretically. A change in its cross
section could determine large variations in the expected surface
barium abundance and, thus, its determination should deserve
renewed efforts.

\begin{figure*}[tpb]
\centering
\includegraphics[width=\textwidth]{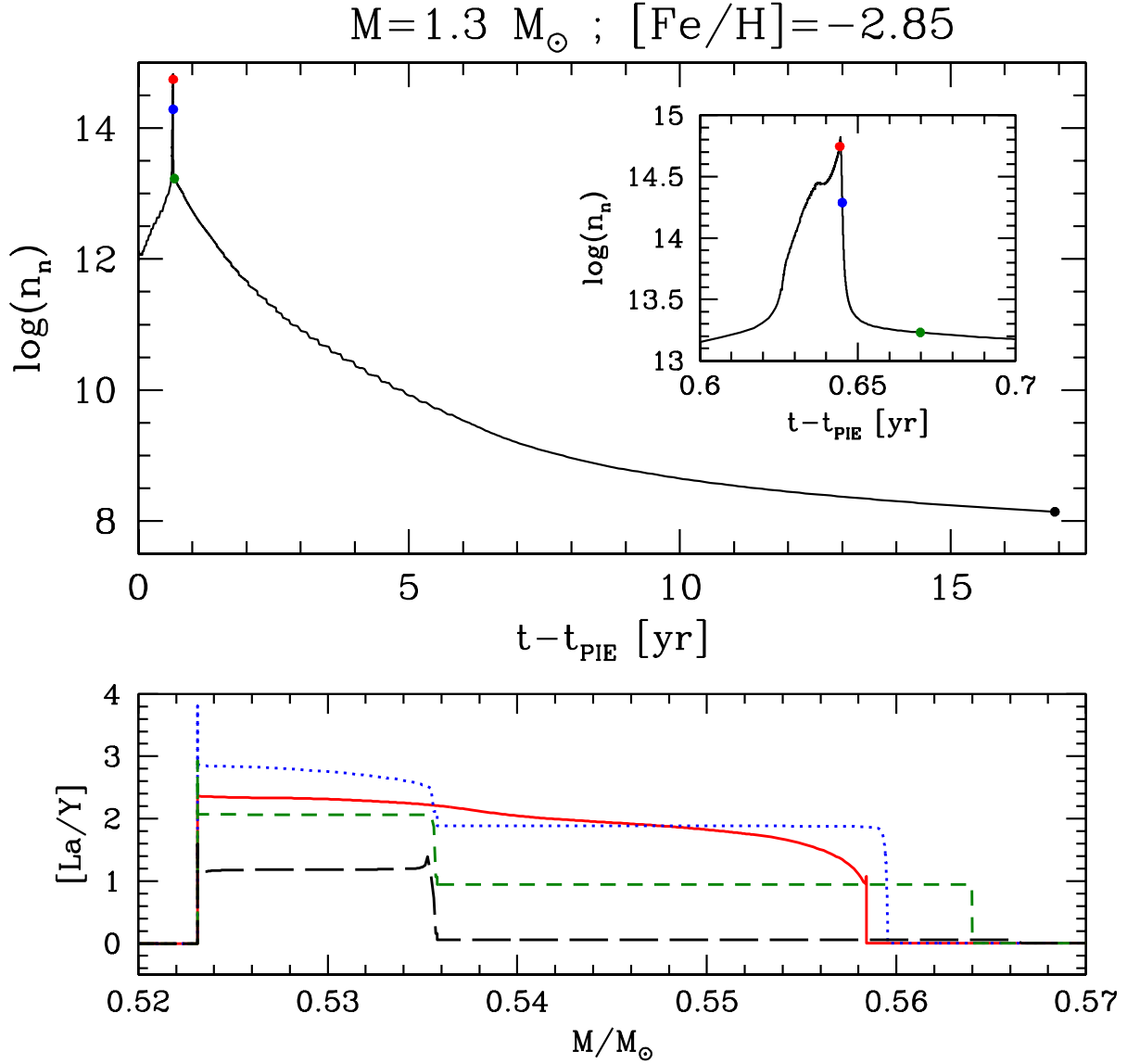}
\caption{Upper panel: maximum neutron density reached during the
PIE of a 1.3 M\odo model with [Fe/H]=-2.85 as a function of time
from the onset of the PIE. Lower panel: s-process indexes in
stellar layers affected by the PIE. Each color refers to a
different delay from the onset of the PIE (see corresponding dots
in the upper panel). See text for details.} \label{fig7}
\end{figure*}

The low metallicity models we compute for this paper present a
major difference with respect to that presented in \cite{cri09b},
because we add an initial $\alpha$-elements enhancement
([$\alpha$/Fe]=0.5). Such an enrichment of $\alpha$ elements,
which is normally observed in halo stars \citep{abiarebolo}, has
strong consequences on the occurrence of PIEs. In fact, the
increased initial oxygen abundance makes the H-shell more
efficient, thus contrasting the occurrence of the PIE episode
itself. As a matter of fact, the only models experiencing PIEs are
the 1.3 M\odo and, to a lesser extent, the 1.5 M\odo with
[Fe/H]=-2.85. This is evident in Fig. \ref{fig5}, in which the 1.3
M\odo curve shows a clear decrease at the smallest metallicity,
thus confirming that, in our models, the occurrence of PIEs leads
to a low [hs/ls]. We verify what happens at larger metallicities
([Fe/H]=-2.15) and we find that a very mild PIE occurs in the 1
M\odo only (crossed dot in Fig. \ref{fig5}), without a clear
nucleosynthetic signature in the envelope of this model. Thus, the
introduction of an initial $\alpha$ enhancement strongly reduces
the number of expected stars experiencing PIEs and, consequently,
weakens the related scenario for the formation of CEMP-rs stars
from low mass, low metallicity AGB stars (see e.g.
\citealt{abate2016}). Another important issue related to PIEs
concerns the use of 1D hydrostatic codes, which necessarily
implies a parametric and space-averaged treatment of convection.
Recent 3D calculations \citep{he11,he14,woodward2015}, although
computed for a different class of stars (Sakurai objects), confirm
the occurrence of the splitting of the convective He-shell, even
if with a temporal delay with respect to 1D hydrostatic models. In
those models, convective mixing is usually modelled as a pure
diffusive process. In our models, we use a time-dependent mixing
scheme derived from an algorithm originally proposed by
\cite{sp80} and subsequently modified by \citet{chi98} and
\citet{stra06}. Moreover, in order to handle in-flight proton
burning, we limit the model temporal step to a fraction (we assume
1/3) of the CNO burning timescale. As a consequence, the splitting
occurs after about 0.65 yr from the starting of the PIE, later
than in the models with diffusive mixing. During that temporal
interval, the \ct (which is mixed down to the bottom of the
convective shell) accumulates enough to generate a substantial
s-process nucleosynthesis. In Fig. \ref{fig7} we report the
temporal evolution of the maximum neutron density we obtain in our
1.3 M\odo model with [Fe/H]=-2.85 (upper panel). We attain a very
high neutron density for a very short period, because the increase
in $n_n$ is interrupted by the splitting of the convective
He-shell. The red dot in Figure \ref{fig7} corresponds to the
moment immediately before the splitting of the shell. At that
time, the [La/Y] (representative of the [hs/ls] s-process index)
within the convective shell is already larger than 2 (see red
curve in the lower panel). After the splitting, it still grows
almost up to 3 in the lower shell (blue dotted curve), due to the
\ct reservoir stored before the splitting. Later, considering the
paucity of iron seeds, lead starts being produced at the expense
of La (green short-dashed curve). Finally, after about 17 years
from the onset of the PIE, a low [La/Y] value characterizes the
lower shell (long-dashed dark curve), while in the upper shell an
almost solar value is found. In summary, during a PIE there are
layers in which a very large [hs/ls] can be attained, but the
following evolution may remove every trace of it. Recently,
\cite{dardelet} and \citet{hampel} compared the results of
one-zone network calculations to observed CEMP-rs stars. They
showed that observed [hs/ls] ratios can be matched with very high
neutron densities ($n_n >$ 10$^{15}$ cm$^{-3}$) lasting for about
0.1 yr. Those calculations are very instructive because they
identify the characteristics of the process able to reproduce
observations. The results they obtain are not so different from
what we showed in Figure \ref{fig7} soon after the occurrence of
the splitting of the convective shell. However, post-processing
calculations do not include the physical drawback stimulated by a
PIE and, therefore, they may miss an important part of the
following physical evolution. On the other side, the parametric
treatment of PIEs in our 1D stellar evolutionary code may lead to
wrong conclusions and the splitting may occur under very different
physical conditions, with important consequences on the following
evolution.

\section{Conclusions} \label{conclu}

In this paper we used a homogeneous sample of CH stars
observations to constrain AGB models. The homogeneity of such a
sample is fundamental in order to exclude discrepancies due to the
use of different analysis techniques and/or atmosphere models. We
mainly concentrate on the s-process index [hs/ls] (i.e. the ratio
between elements belonging to the first and to the second peak of
the s-process) because this quantity is almost independent of the
number of experienced TDUs. On average, observations show an
increasing [hs/ls] with decreasing initial metal content as well
as a consistent spread for a fixed metallicity. Theoretical models
described here have been calculated with the FUNS code and are
available on the web pages of the FRUITY database. We find that
our standard set reproduces the increase in the [hs/ls] down to
[Fe/H]=-1 fairly well. However, when taking into consideration the
expected dilution due to the binary origin of CH stars, models are
not able to cover the whole observational sample. Such a
disagreement can be improved by adopting a different criterion for
handling the radiative/convective interface at the base of the
convective envelope as well as taking explicitly into
consideration the effects induced by rotation. Notwithstanding,
even when including the aforementioned physical processes in our
stellar evolutionary code, we cannot match the high [hs/ls] ratios
observed in low metallicity CH and CEMP stars. We plan to evaluate
the effects induced by other physical mechanisms on the ongoing
s-process nucleosynthesis, such as mixing induced by magnetic
fields. Finally, we check the possibility that proton ingestion
episodes occurring at the beginning of the AGB phase of low mass
stars could solve the problem. In our models, even if a PIE
episode leads to a provisional and very high local [hs/ls] ratio,
the following evolution removes any trace of it and, thus, the
surface is characterized by a low [hs/ls] ratio. Moreover, the use
of an initial mixture that is enriched in $\alpha$-elements limits
the occurrence of PIEs to low mass models (M$\le$ 1.3 M$_\odot$)
at very low metallicities ([Fe/H]$\le$-2.85).

\section {Acknowledgement}

We thank the anonymous referee for valuable comments on the paper.
SC, LP and DG acknowledge funding from by Italian grant PRIN-MIUR
2012 ``Nucleosynthesis in AGB stars: An integrated approach''
project (20128PCN59). AG gratefully acknowledges funding from the
DST project SB/S2/HEP-010/2013. We are indebted with M. Quintini
for his precious help in managing and updating the FRUITY database
web platform. We acknowledge M. Eichler for a careful reading of
the manuscript. S.C. thanks S.W. Campbell for stimulating
scientific discussions which influenced this paper.

\bibliographystyle{aa}
\bibliography{paper_CH_stars}

\end{document}